\documentclass[11pt]{amsart}

\usepackage{amssymb,latexsym}
\usepackage{graphicx,epsfig,color}

\newtheorem{thm}{Theorem}[section]

\newtheorem{lem}[thm]{Lemma}

\newtheorem{defn}[thm]{Definition}

\numberwithin{equation}{section}

\date{}

\begin{document}

\author[Tulkin H. Rasulov and Elyor B. Dilmurodov]{Tulkin H. Rasulov and Elyor B. Dilmurodov}
\title[Analysis of the spectrum of a $2\times 2$ operator matrix]
{Analysis of the spectrum of a $2\times 2$ operator matrix. Discrete spectrum asymptotics} \maketitle

\begin{center}
{\small Faculty of Physics and Mathematics, Bukhara State University\\
M. Ikbol str. 11, 200100 Bukhara, Uzbekistan\\
E-mail: rth@mail.ru, elyor.dilmurodov@mail.ru}
\end{center}

\begin{abstract}
We consider a $2 \times 2$ operator matrix
${\mathcal A}_\mu,$ $\mu>0$ related with the
lattice systems describing two identical bosons and one particle,
another nature in interactions, without conservation of the number
of particles. We obtain an analogue of the Faddeev
equation and its symmetric version for the eigenfunctions of ${\mathcal A}_\mu$. We describe the
new branches of the essential spectrum of
${\mathcal A}_\mu$ via the spectrum of a family of generalized Friedrichs models.
It is established that the essential spectrum of ${\mathcal A}_\mu$
consists the union of at most three bounded closed intervals and their location is studied.
For the critical value $\mu_0$ of the coupling constant $\mu$ we
establish the existence of infinitely many eigenvalues, which are located in the both sides of
the essential spectrum of ${\mathcal A}_\mu$. In this case, an asymptotic formula for the discrete
spectrum of ${\mathcal A}_\mu$ is found.
\end{abstract}



\section{Introduction and statement of the problem}

It is well-known that \cite{CT08}, if $H$ is a bounded linear operator in a Hilbert space ${\mathcal H}$
and a decomposition ${\mathcal H}={\mathcal H}_1 \oplus {\mathcal H}_2$ into two Hilbert spaces
${\mathcal H}_1$, ${\mathcal H}_2$ is given, then $H$ always admits a block operator matrix
representation
\begin{equation*}
H=\left(
\begin{array}{cc}
H_{11} & H_{12}\\
H_{21} & H_{22}\\
\end{array}
\right)
\end{equation*}
with bounded linear operators $H_{ij}: {\mathcal H}_j \to {\mathcal H}_i$, $i,j=1,2$.
In addition, $H=H^*$ if and only if $H_{ii}=H_{ii}^*$, $i=1,2$ and $H_{21}=H_{12}^*$.
Such operator matrices often arise in mathematical physics, e.g. in quantum
field theory, condensed matter physics, fluid mechanics,
magnetohydrodynamics and quantum mechanics. One of the special class
of $2 \times 2$ block operator matrices is the Hamiltonians acting in the one- and two-particle
subspaces of a Fock space. It is related with a system
describing three-particles in interaction without conservation of the
number of particles in Fock space. Here off-diagonal entries of such
block operator matrices are annihilation and creation operators.

Operator matrices of this form play a key role for the study of the energy operator of the spin-boson Hamiltonian
with two bosons on the torus. In fact, the latter is a $6 \times 6$ operator matrix which is unitarily
equivalent to a $2 \times 2$ block diagonal operator with two copies of a particular case of $H$ on the diagonal,
see e.g. \cite{MNR}. Consequently, the location of the essential spectrum and finiteness of discrete
eigenvalues of the spin-boson Hamiltonian are determined by the corresponding spectral information on the
operator matrix $H$. We recall that the spin-boson model is a well-known quantum-mechanical model which describes
the interaction between a two-level atom and a photon field. We refer to \cite{LChDFGZ} and \cite{HSp95} for excellent reviews from physical and mathematical perspectives, respectively.
Independently of whether the underlying domain is a torus ${\Bbb T}^{\rm d}$ or the whole space ${\Bbb R}^{\rm d}$,
the full spin-boson Hamiltonian is an infinite operator matrix in Fock space for which rigorous results are very hard to obtain. One line of attack is to consider the compression to the truncated Fock space with a finite number $N$ of bosons,
and in fact most of the existing literature concentrates on the case $N \leq 2$. For the case of ${\Bbb R}^{\rm d}$
there are some exceptions, e.g. \cite{Gerard96}, \cite{HSp95-1} for arbitrary finite $N$ and \cite{ZhM95} for $N=3$, where a rigorous scattering theory
was developed for small coupling constants. In \cite{OI2018} it is shown that the discrete spectrum
of the spin-boson model with two photons in ${\Bbb R}^{\rm d}$ is finite and 
the essential spectrum consists of a half-line, the bottom of which is a unique zero of a simple Nevanlinna function.

For the case when the underlying domain is a torus, the spectral properties of some versions of $H$
were investigated in \cite{ALR}, \cite{ALR1}, \cite{OT2018}, \cite{MR14}, \cite{RT2019}.
An important problem of the spectral theory of such matrix operators is the infiniteness
of the number of eigenvalues located outside the essential spectrum.
We mention that, the infiniteness of the discrete eigenvalues below the bottom
of the essential spectrum of the Hamiltonian in Fock space,
which has a block operator matrix representation, and corresponding discrete spectrum asymptotics were
discussed in \cite{ALR}, \cite{Ras2011}. These results were obtained using the machinery developed in \cite{Sob}
by Sobolev.

In the present paper we consider a $2 \times 2$
operator matrix ${\mathcal A}_\mu,$ ($\mu>0$ is a coupling constant) related with
the lattice systems describing two identical bosons and one
particle, another nature in interactions, without conservation of
the number of particles. This operator acts in the direct sum of
one- and two-particle subspaces of the bosonic Fock space
and it is related with the lattice spin-boson Hamiltonian
\cite{MNR, Ras2016}. We find the critical value $\mu_0$ of the coupling constant $\mu$,
to establish the existence of infinitely many eigenvalues lying in
{\bf both} sides of essential spectrum of ${\mathcal A}_{\mu_0}$ and to obtain an asymptotics for the
number of these eigenvalues.

We point out that the latter assertion seems to be quite new
for the discrete models and similar result have not been obtained yet for the
three-particle discrete Schr\"{o}dinger operators and
operator matrices in Fock space.
In all papers devoted to the infiniteness of the number of eigenvalues (Efimov's effects)
the situation on the neighborhood of the left edge of essential spectrum are discussed,
see for example \cite{AL, ALM, ALR, ALR1, LM, MR14}.
Since the essential spectrum of
the three-particle continuous Schr\"{o}dinger operators \cite{Mog, RS4, Sob} and standard spin-boson model with
at most two photons \cite{Mal-Min, MS} coincides with half-axis $[\kappa; +\infty)$, the main results of
the present paper are typical only for lattice case, and they do not have analogues in the continues case.

Now we formulate the problem.
Let ${\Bbb T}^3$ be the three-dimensional torus, the cube
$(-\pi,\pi]^3$ with appropriately identified sides equipped with
its Haar measure. Let $L_2({\Bbb T}^3)$
be the Hilbert space of square integrable (complex) functions
defined on ${\Bbb T}^3$ and $ L_2^{\rm s}(({\Bbb T}^3)^2)$ be the
Hilbert space of square integrable (complex) symmetric functions
defined on $({\Bbb T}^3)^2.$ Denote by ${\mathcal H}$ the direct
sum of spaces ${\mathcal H}_1:=L_2({\Bbb
T}^3)$ and ${\mathcal H}_2:=L_2^{\rm s}(({\Bbb T}^3)^2),$ that is,
${\mathcal H}:={\mathcal H}_1 \oplus {\mathcal H}_2.$ The spaces ${\mathcal H}_1$ and
${\mathcal H}_2$ are called one- and two-particle subspaces
of a bosonic Fock space ${\mathcal F}_{\rm s}(L_2({\Bbb T}^3))$
over $L_2({\Bbb T}^3),$ respectively.

Let us consider a $2 \times 2$ operator
matrix ${\mathcal A}_\mu$ acting in the Hilbert
space ${\mathcal H}$ as
$$
{\mathcal A}_\mu:=\left( \begin{array}{cc}
A_{11} & \mu A_{12}\\
\mu A_{12}^* & A_{22}\\
\end{array}
\right)
$$
with the entries
\begin{align*}
& (A_{11}f_1)(k)=w_1(k)f_1(k), \quad
(A_{12}f_2)(k)= \int_{{\Bbb T}^3} f_2(k,s)ds,\\
& (A_{22}f_2)(k,p)=w_2(k,p)f_2(k,p),\quad f_i \in {\mathcal H}_i,\quad i=1,2.
\end{align*}
Here $\mu>0$ is a coupling constant, the functions
$w_1(\cdot)$ and $w_2(\cdot, \cdot)$ have the form
$$
w_1(k):=\varepsilon(k)+\gamma, \quad
w_2(k,p):=\varepsilon(k)+\varepsilon(\frac{1}{2}(k+p))+\varepsilon(p)
$$
with $\gamma \in {\Bbb R}$ and the dispersion function $\varepsilon(\cdot)$
is defined by
\begin{equation}\label{epsilon}
\varepsilon(k):=\sum_{i=1}^3 (1-\cos \, k_i),\,k=(k_1, k_2, k_3) \in
{\Bbb T}^3,
\end{equation}
$A_{12}^*$  denotes the adjoint operator to $A_{12}$
and
$$
(A_{12}^*f_1)(k,p)=\frac{1}{2} (f_1(k)+f_1(p)), \quad f_1 \in {\mathcal H}_1.
$$

Under these assumptions the operator ${\mathcal A}_\mu$ is bounded
and self-adjoint.

We remark that the operators $A_{12}$ and $A_{12}^*$ are called
annihilation and creation operators \cite{Frid}, respectively.
In physics, an annihilation operator is an operator that lowers
the number of particles in a given state by one, a creation
operator is an operator that increases the number of particles in
a given state by one, and it is the adjoint of the annihilation
operator.

\section{Faddeev's equation and essential spectrum of ${\mathcal A}_\mu$}

In this section we obtain an analogue of the Faddeev type integral equation for eigenvectors of ${\mathcal A}_\mu$
and investigate the location and structure of the essential spectrum of ${\mathcal A}_\mu$.

Throughout the present paper we adopt the following conventions: Denote by $\sigma(\cdot),$ $\sigma_{\rm
ess}(\cdot)$ and $\sigma_{\rm disc}(\cdot),$ respectively, the
spectrum, the essential spectrum, and the discrete spectrum of a
bounded self-adjoint operator.

Let $H_0:={\Bbb C}$. To study the spectral properties of the operator ${\mathcal A}_\mu$ we
introduce a family of bounded self-adjoint operators (generalized Friedrichs
models) ${\mathcal A}_\mu(k),$ $k\in {\Bbb T}^3$ which acts in
${\mathcal H}_0 \oplus {\mathcal H}_1$ as $2 \times 2$ operator matrices
$$
{\mathcal A}_\mu(k):=\left( \begin{array}{cc}
A_{00}(k) & \frac{\mu}{\sqrt{2}} A_{01}\\
\frac{\mu}{\sqrt{2}} A_{01}^* & A_{11}(k)\\
\end{array}
\right),
$$
with matrix elements
\begin{align*}
&  A_{00}(k)f_0=w_1(k) f_0,\,\,
(A_{01}f_1)= \int_{{\Bbb T}^3} f_1(t)dt,\\
& (A_{11}(k)f_2)(p)=w_2(k,p)f_1(p), \quad f_i \in {\mathcal H}_i, \quad i=1,2.
\end{align*}

From the simple discussions it follows that $\sigma_{\rm ess}({\mathcal A}_\mu (k))=[m(k), M(k)],$
where the numbers $m(k)$ and $M(k)$ are defined by
\begin{equation}\label{m(p) and M(p)}
m(k):=\min\limits_{p\in {\Bbb T}^3} w_2(k,p), \quad M(k):=
\max\limits_{p\in {\Bbb T}^3} w_2(k,p).
\end{equation}

For any $k\in {\Bbb T}^3$ we define an analytic function
$I(k\,; \cdot)$ in ${\Bbb C} \setminus
\sigma_{\rm ess}({\mathcal A}_\mu(k))$ by
$$
I(k\,; z):=\int_{{\Bbb T}^3} \frac{dt}{w_2(k,t)-z}.
$$
Then the Fredholm determinant associated to
the operator ${\mathcal A}_\mu(k)$ is defined by
\begin{equation*}
\Delta_\mu(k\,; z):=w_1(k)-z-\frac{\mu^2}{2}\, I(k\,; z),\,\, z\in {{\Bbb C} \setminus
\sigma_{\rm ess}({\mathcal A}_\mu(k))}.
\end{equation*}

A simple consequence of the Birman-Schwinger principle and the Fredholm theorem implies
that for the discrete spectrum of ${\mathcal A}_\mu (k)$ the equality
$$
\sigma_{\rm disc}({\mathcal A}_\mu (k))=\{z \in {\Bbb C} \setminus [m(k); M(k)]:\,
\Delta_\mu(k\,; z)=0 \}
$$
holds.

Set
\begin{align*}
& m:= \min\limits_{k,p\in {\Bbb T}^3} w_2(k,p),
\quad M:= \max\limits_{k,p\in {\Bbb T}^3} w_2(k,p), \\
& \Lambda_\mu:=\bigcup_{k \in {\Bbb T}^3}
\sigma_{\rm disc}({\mathcal A}_\mu (k)), \quad \Sigma_\mu:=[m; M] \cup \Lambda_\mu.
\end{align*}

For each $\mu>0$ and $z \in {\Bbb C} \setminus \Sigma_\mu$ we define the integral operator $T_\mu(z)$ acting in the Hilbert spaces $L_2({\Bbb T}^3)$ by
$$
(T_\mu(z)g)(p)=\frac{\mu^2 }{2\Delta_\mu(p;\, z)}\int_{{\Bbb T}^3}
\frac{g(t)dt}{w_2(p,t)-z}.
$$

The following theorem is an analog of the well-known Faddeev's
result for the operator ${\mathcal A}_\mu$ and establishes a connection between
eigenvalues of ${\mathcal A}_\mu$ and $T_\mu(z).$

\begin{thm}\label{Main Theorem 1}
The number $z \in {\Bbb C} \setminus \Sigma_\mu$
is an eigenvalue of the operator ${\mathcal A}_\mu$ if and only if the number
$\lambda=1$ is an eigenvalue of the operator $T_\mu(z).$
Moreover the eigenvalues $z$ and $1$ have the same multiplicities.
\end{thm}

We point out that the integral equation $g=T_\mu(z) g$
is an analogue of the Faddeev type system of integral equations for eigenfunctions of the
operator ${\mathcal A}_\mu$ and it is played crucial role in the analysis of the
spectrum of ${\mathcal A}_\mu.$ For the proof of Theorem \ref{Main Theorem 1} we show the equivalence
of the eigenvalue problem ${\mathcal A}_\mu f=zf$ to the equation $g=T_\mu(z) g$.

The following theorem describes the location of
the essential spectrum of the operator ${\mathcal A}_\mu$ by the
spectrum of the family of generalized Friedrichs models ${\mathcal A}_\mu (k)$.

\begin{thm}\label{Main Theorem 2} For the essential spectrum of ${\mathcal A}_\mu$ the
equality $\sigma_{\rm ess}({\mathcal A}_\mu)=\Sigma_\mu$ holds.
Moreover the set $\Sigma_\mu$ consists of no more than three
bounded closed intervals.
\end{thm}

The inclusion $\Sigma_\mu \subset \sigma_{\rm ess}({\mathcal A}_\mu)$ in the proof
of Theorem \ref{Main Theorem 2} is established with the use of a well-known
Weyl creterion, see for example \cite{RT2019}. An application of Theorem \ref{Main Theorem 1} and analytic
Fredholm theorem (see, e.g., Theorem VI.14 in \cite{RS4}) proves inclusion
$\sigma_{\rm ess}({\mathcal A}_\mu) \subset \Sigma_\mu.$

In the following we introduce the new subsets of the essential spectrum of ${\mathcal A}_\mu.$

\begin{defn}
The sets $\Lambda_\mu$ and $[m; M]$ are called two- and three-particle
branches of the essential spectrum of ${\mathcal A}_\mu,$ respectively.
\end{defn}

The definition of the set $\Lambda_\mu$ and the equality
$$
\bigcup_{k \in {\Bbb T}^3} [m(k); M(k)]=[m; M]
$$
together with Theorem \ref{Main Theorem 2} give the equality
\begin{equation}\label{essential spectrum of mathcal A}
\sigma_{\rm ess}({\mathcal A}_\mu)=\bigcup_{k \in {\Bbb T}^3} \sigma({\mathcal A}_\mu (k)).
\end{equation}
Here the family of operators ${\mathcal A}_\mu (k)$ have a simpler structure than the operator ${\mathcal A}_\mu.$
Hence, in many instance, \eqref{essential spectrum of mathcal A}
provides an effective tool for the description
of the essential spectrum.

Using the extremal properties of the function $w_2(\cdot,\cdot),$
and the Lebesgue dominated convergence theorem one can show that
the integral $I(\overline{0};0)$ is finite, where $\bar{0}:=(0,0,0) \in {\Bbb T}^3$, see \cite{RD2019, RD2019-1}.

For the next investigations we introduce the following quantities
\begin{align*}
& \mu_l^0(\gamma):=\sqrt{2\gamma} \left( I(\overline{0},0) \right)^{-1/2}\,\, \mbox{for}\,\, \gamma>0;\\
& \mu_r^0(\gamma):=\sqrt{24-2\gamma} \left( I(\overline{0},0)
\right)^{-1/2}\,\, \mbox{for}\,\, \gamma<12.
\end{align*}

Since ${{\Bbb T}^3}$ is compact, and the functions $\Delta_\mu(\cdot;0)$ and $\Delta_\mu(\cdot;18)$ are
continuous on ${{\Bbb T}^3}$, there exist points
$k_0,k_1\in {{\Bbb T}^3}$ such that the equalities
\begin{align*}
\max\limits_{k\in {\Bbb T}^3} \Delta_\mu(k;0)=\Delta_\mu(k_0;0),
\quad \min\limits_{k\in {\Bbb T}^3}
\Delta_\mu(k;18)=\Delta_\mu(k_1;18)
\end{align*}
hold.

Let us define the following notations:
\begin{align*}
 & \gamma_0:=\left(
12\frac{I(k_0;0)}{I(\overline{0};0)}-\varepsilon(k_0)
\right) \left( 1+\frac{I(k_0;0)}{I(\overline{0};0)}\right)^{-1};\\
& \gamma_1:=\left(18-\varepsilon(k_1)\right) \left(
1-\frac{I(k_1;18)}{I(\overline{0};0)}\right).
\end{align*}

Denote
\begin{align*}
& E_{\mu}^{(1)}:=\text {min}\left\{\Lambda_\mu\cap(-\infty;0]
\right\};
E_\mu^{(2)}:=\text {max}\left\{ \Lambda_\mu\cap(-\infty;0] \right\};\\
& E_{\mu}^{(3)}:=\text {min}\left\{
\Lambda_\mu\cap[18;\infty)\right\};
E_\mu^{(4)}:=\text {max}\left\{ \Lambda_\mu\cap[18;\infty)\right\}.\\
\end{align*}

We formulate the results, which
are precisely describe the structure of the essential spectrum of
${\mathcal A}_\mu$. The structure of the essential spectrum depends
on the location of the parameters $\mu>0$ and $\gamma \in {\Bbb R}$.

\begin{thm}\label{THM3.1}
Let $\mu=\mu_r^0(\gamma),$ with
$\gamma<12.$  The following equality holds
$$
\sigma_{\rm ess}({\mathcal A}_\mu)=\begin{cases}
[E_1;E_2]\bigcup[0;18], & \text {if $ \gamma < \gamma_0 $}; \\
[E_1;18], & \text {if $ \gamma_0 \leq \gamma < 6$;}\\
[0;18], & \text {if $ 6 \leq \gamma<12 $.}
\end{cases}
$$
\end{thm}

\begin{thm}\label{THM3.2}
Let $\mu=\mu_l^0(\gamma),$ with $\gamma>0.$
 The following equality holds
$$
\sigma_{\rm ess}({\mathcal A}_\mu)=\begin{cases}
[0;18], & \text{if $ 0<\gamma \leq 6$}; \\
[0;E_\mu^{(4)}], & \text{if $ 6 < \gamma \leq \gamma_1 $};\\
[0;18]\bigcup[E_\mu^{(3)};E_\mu^{(4)}], & \text {if $ \gamma > \gamma_1$.}
\end{cases}
$$
\end{thm}

The proof of these two theorems are based on the existence conditions of the eigenvalue $z_\mu(k)$
of the operator ${\mathcal A}_\mu(\cdot)$ and the continuity of $z_\mu(\cdot)$ on its domain.

\section{Birman-Schwinger principle and discrete spectrum asymptotics of the operator ${\mathcal A}_\mu$}

Let us denote by $\tau_{\min}({\mathcal A}_\mu)$ and $\tau_{\max}({\mathcal A}_\mu)$ the lower and upper bounds of the essential spectrum $\sigma_{\rm ess}({\mathcal A}_\mu)$ of the operator ${\mathcal A}_\mu$, respectively, that is,
$$
\tau_{\min}({\mathcal A}_\mu):\equiv \min\sigma_{\rm ess}({\mathcal A}_\mu), \quad
\tau_{\max}({\mathcal A}_\mu):\equiv \max\sigma_{\rm ess}({\mathcal A}_\mu).
$$

For an interval $\Delta \subset {\Bbb R},$ $E_{\Delta}({\mathcal A}_\mu)$
stands for the spectral subspace of ${\mathcal A}_\mu$ corresponding to $\Delta.$
Let us denote by $\sharp\{\cdot\}$ the cardinality of a set and
by $N_{(a, b)}({\mathcal A}_\mu)$ the number of eigenvalues of the operator ${\mathcal A}_\mu,$
including multiplicities, lying in $(a, b) \subset {\Bbb R}
\setminus \sigma_{\rm ess}({\mathcal A}_\mu),$ that is,
$$
N_{(a, b)}({\mathcal A}_\mu):=\dim E_{(a, b)}({\mathcal A}_\mu).
$$

For a $\lambda \in {\Bbb R},$ we define the number $n(\lambda, A_\mu)$ as
follows
$$
n(\lambda, A_\mu):=\sup \{{\rm dim} F: (A_\mu u,u)>\lambda,\, u\in F \subset
{\mathcal H},\,||u||=1\}.
$$

The number $n(\lambda, A_\mu)$ is equal to the infinity if
$\lambda<\max\sigma_{\rm ess}(A_\mu);$ if $n(\lambda, A_\mu)$ is finite,
then it is equal to the number of the eigenvalues of $A_\mu$ bigger
than $\lambda.$

By the definition of $N_{(a; b)}({\mathcal A}_\mu),$ we have
\begin{align*}
N_{(-\infty; z)}({\mathcal A}_\mu)&=n(-z, -{\mathcal A}_\mu),\,-z>-\tau_{\min}({\mathcal A}_\mu),\\
N_{(z; +\infty)}({\mathcal A}_\mu)&=n(z, {\mathcal A}_\mu),\,z>\tau_{\max}({\mathcal A}_\mu).
\end{align*}

In our analysis of the discrete spectrum of ${\mathcal A}_\mu$ the crucial role is played by the compact operator
$\widehat{T}_\mu(z)$, $z\in {\Bbb R}\setminus [\tau_{\min}({\mathcal A}_\mu); \tau_{\max}({\mathcal A}_\mu)]$ in the space $L_2({\Bbb T}^3)$ as integral operator
\begin{align*}
(\widehat{T}_\mu(z)g)(p)&=\frac{\mu^2}{2\sqrt{\Delta_\mu(p;\,z)}}\int_{{\Bbb T}^3} \frac{g(t)dt}{\sqrt{\Delta_\mu(t;\,z)}(w_2(p,t)-z)},
\quad \mbox{for} \quad z<\tau_{\min}({\mathcal A}_\mu),\\
(\widehat{T}_\mu(z)g)(p)&=-\frac{\mu^2}{2\sqrt{-\Delta_\mu(p;\,z)}}\int_{{\Bbb T}^3} \frac{g(t)dt}{\sqrt{-\Delta_\mu(t;\,z)}(w_2(p,t)-z)},
\quad \mbox{for} \quad z>\tau_{\max}({\mathcal A}_\mu).
\end{align*}

The following lemma is a realization of the well-known
Birman-Schwinger principle for the operator ${\mathcal A}_\mu$ (see \cite{ALR}).

\begin{lem}\label{LEM 4.3} For $z\in {\Bbb R}\setminus [\tau_{\min}({\mathcal A}_\mu); \tau_{\max}({\mathcal A}_\mu)]$ the
operator $\widehat{T}_\mu(z)$ is compact and continuous in
$z$ and
\begin{align*}
N_{(-\infty; z)}({\mathcal A}_\mu)& = n(1, \widehat{T}_\mu(z))\quad \mbox{for}\quad z<\tau_{\min}({\mathcal A}_\mu),\\
N_{(z; +\infty)}({\mathcal A}_\mu)&=n(1, \widehat{T}_\mu(z))\quad \mbox{for}\quad z>\tau_{\max}({\mathcal A}_\mu).
\end{align*}
\end{lem}

This lemma can be proven quite similarly to the corresponding result of \cite{ALR}.

Let ${\Bbb S}^2$ being the unit sphere in ${\Bbb R}^3$ and
$$
S_r: L_2((0, r), \sigma_0) \to L_2((0, r), \sigma_0), \quad
r>0, \quad \sigma_0=L_2({\Bbb S}^2)
$$
be the integral operator with the kernel
$$
S(t; y)=\frac{25}{8\pi^2 \sqrt{6}}\, \frac{1}{5 \cos(h y)+t},
$$
$$
y=x-x', \quad x, x' \in (0, r), \quad t=(\xi, \eta), \quad \xi,
\eta \in {\Bbb S}^2.
$$

For $\lambda>0,$ define
$$
U(\lambda)=\frac{1}{2} \lim_{r \to \infty} r^{-1} n(\lambda, S_r).
$$

The existence of the latter limit and the fact $U(1)>0$ shown in
\cite{Sob}.

From the definitions of the quantities $\mu_l^0(\gamma)$ and $\mu_r^0(\gamma)$ it is easy to see that
$\mu_l^0(6)=\mu_r^0(6)$. We set $\mu_0:=\mu_l^0(6)$.

We can now formulate our last main result.

\begin{thm}\label{main theorem-3}
The following relations hold:
$$
\sharp (\sigma_{\rm disc} ({\mathcal A}_{\mu_0}) \cap (-\infty, 0))=
\sharp (\sigma_{\rm disc} ({\mathcal A}_{\mu_0}) \cap (18,
\infty))=\infty;
$$
\begin{equation}\label{main asymp}
\lim\limits_{z \nearrow 0}\frac{N_{(-\infty,\,
z)}({\mathcal A}_{\mu_0})}{|\log|z||}= \lim\limits_{z \searrow 18}
\frac{N_{(z,\, \infty)}({\mathcal A}_{\mu_0})}{|\log|z-18||}=U(1).
\end{equation}
\end{thm}

Clearly, by equality \eqref{main asymp} the infinite cardinality of the
parts of discrete spectrum of ${\mathcal A}_{\mu_0}$ in $(-\infty; 0)$ and $(18; +\infty)$ follows automatically
from the positivity of $U(1)$.

\end{document}